\documentclass[journal,transmag]{IEEEtran}

\usepackage{graphicx}
\graphicspath{{./}{./images/}{./images_all/}{./images_all/scaling/}}
\DeclareGraphicsExtensions{.eps}

\usepackage[cmex10]{amsmath}

\usepackage{subfigure}

\begin{document}
\title{Area-Delay-Energy Tradeoffs of Strain-Mediated\\ Multiferroic Devices}

\author{\IEEEauthorblockN{Kuntal Roy}
\IEEEauthorblockA{School of Electrical and Computer Engineering, Purdue University, West Lafayette, IN 47907, USA}
\thanks{Contact: Kuntal Roy (email: royk@purdue.edu).}
\thanks{Some of the work for this paper was performed prior to K. Roy joining Purdue University.}}

\markboth{Area-Delay-Energy Tradeoffs of Strain-Mediated Multiferroic Devices}
{Roy, K.}

\IEEEtitleabstractindextext{%
\begin{abstract}
Multiferroic devices hold profound promise for ultra-low energy computing in beyond Moore's law era. The magnetization of a magnetostrictive shape-anisotropic single-domain nanomagnet strain-coupled with a piezoelectric layer in a multiferroic composite structure can be switched between its two stable states (separated by an energy barrier) with a tiny amount of voltage via converse magnetoelectric effect. With appropriate choice of materials, the magnetization can be switched with a few tens of millivolts of voltages in sub-nanosecond switching delay while spending a miniscule amount of energy of $\sim$1 attojoule at room-temperature. Here, we analyze the area-delay-energy trade-offs of these multiferroic devices by solving stochastic Landau-Lifshitz-Gilbert equation in the presence of room-temperature thermal fluctuations. We particularly put attention on scaling down the lateral area of the magnetostrictive nanomagnet that can increase the device density on a chip. We show that the vertical thickness of the nanomagnet can be increased while scaling down the lateral area and keeping the assumption of single-domain limit valid. This has important consequence since it helps to some extent preventing the deterioration of the induced stress-anisotropy energy in the magnetostrictive nanomagnet, which is proportional to the nanomagnet's volume. The results show that if we scale down the lateral area, the switching delay increases while energy dissipation decreases. Avenues available to decrease the switching delay while still reducing the energy dissipation are discussed.
\end{abstract}

}

\maketitle


\section{Introduction}
\label{sec:introduction}

\begin{figure}
\centering
\includegraphics[width=3.4in]{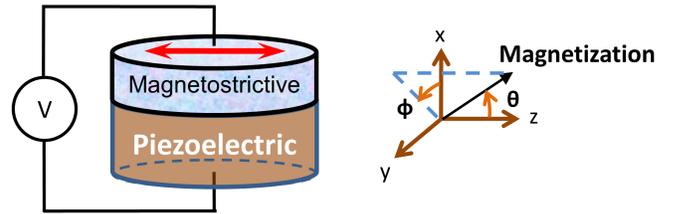}%
\caption{A voltage-controlled strain-mediated multiferroic composite device. The magnetization of the single-domain magnetostrictive nanomagnet can be switched between its two stable states by the generated stress on the nanomagnet via applying a voltage across the composite structure. The voltage generates strain in the piezoelectric layer, which is transferred elastically to the magnetostrictive layer.}
\label{fig:multiferroic_heterostructure}
\end{figure}

\IEEEPARstart{S}{train-mediated} multiferroic devices, i.e., a magnetostrictive layer strain-coupled to a piezoelectric layer works according to the principles of converse magnetoelectric effect~\cite{roy13_spin,roy11,RefWorks:558,Refworks:164,Refworks:165}. When a voltage is applied across such heterostructure, the piezoelectric layer gets strained and the strain is transferred elastically to the magnetostrictive layer and generates a stress-anisotropy (or magnetoelastic anisotropy) in it. If we consider a shape-anisotropic single-domain nanomagnet having two stable magnetization states as the magnetostrictive layer, the generated stress-anisotropy can overcome the shape-anisotropy energy barrier and rotate the magnetization. (See Fig.~\ref{fig:multiferroic_heterostructure}.) With appropriate choice of materials, the magnetization can be switched between its two stable states separated by an energy barrier in sub-nanosecond switching delay while expending miniscule amount of energy of $\sim$1 attojoule at room-temperature~\cite{roy11,roy11_6}. Such study has opened up a new field called \emph{straintronics}~\cite{roy13_spin,roy13,roy14}, which can possibly replace the conventional charge-based electronics as our future information processing paradigm. Experimental efforts demonstrating such electric field induced strain-mediated magnetization rotation are emerging too~\cite{RefWorks:806,RefWorks:609,RefWorks:825,RefWorks:790,RefWorks:824}. It should be emphasized that remedies for substrate clamping effect particularly for low-thickness piezoelectric layers ($<$ 100 nm) need to be considered~\cite{RefWorks:823,RefWorks:820}. The use of thin films rather than thick  substrates~\cite{RefWorks:143} for piezoelectric layers allows us to work with lower voltages and therefore it decreases the energy dissipation.

The stress-anisotropy induced in the magnetostrictive nanomagnet is proportional to the magnetostrictive coefficient of the material used, the generated stress (which is in turn proportional to the electric field), and the volume of the nanomagnet. For a given material with a certain magnetostrictive coefficient and a fixed volume, a higher stress facilitates having a faster switching speed (for a sufficiently fast ramp rate of stress) but causes a higher energy dissipation~\cite{roy11_6}. The energy dissipation in these multiferroic devices comprises of two components: one is due to magnetization damping through which magnetization relaxes to a minimum energy position dissipating energy and the energy dissipation in the external circuitry due to applied voltage ($\propto CV^2$, where $C$ is the capacitance of the piezoelectric layer and $V$ is the applied voltage)~\cite{roy11,roy11_6}. Since the voltage required to switch the magnetization is miniscule, the ``$CV^2$'' energy dissipation is also miniscule.

In this paper, we study the area-delay-energy trade-offs of such multiferroic magnetoelectric devices. We particularly analyze the effect of reducing lateral area to scale down the area consumption on a chip. Scaling down lateral area for a constant thickness reduces both the shape-anisotropic energy barrier height separating the two stable magnetization states and the generated stress-anisotropy in the nanomagnet since they are both proportional to the volume of the nanomagnet. Apparently, the concern is how to deal with such scenario if we wish to decrease the lateral area of such devices. It turns out that while we reduce the lateral area, we can increase the vertical thickness of the nanomagnet to some extent keeping the single-domain limit assumption for the nanomagnet intact. Increasing thickness while reducing lateral area of the nanomagnet facilitates preventing a drastic reduction in volume and hence the induced stress-anisotropy in the nanomagnet does not deteriorate much. Note that we need to keep the shape-anisotropic energy barrier constant since that determines the error-probability of spontaneous reversal of magnetization between two stable states. The decrease in shape-anisotropic energy due to decrease in volume needs to be compensated by properly adjusting the lateral dimensions. We perform simulations using stochastic Landau-Lifshitz-Gilbert (LLG) equation of magnetization dynamics~\cite{RefWorks:162,RefWorks:161,RefWorks:186} to extract the switching delay and energy dissipation during magnetization reversal.



If we increase the thickness of the nanomagnet, the thickness of the piezoelectric layer needs to be also proportionately increased since the effectiveness of strain-transfer depends on the ratio of the two layers~\cite{roy11}. Modeling the piezoelectric layer as a parallel-plate capacitor and noting that the area, $A$ of the capacitor decreases while the thickness $d$ increases, the capacitance of the piezoelectric layer ($\propto A/d$) decreases significantly. The lateral area decreases more compared to the increase in thickness of the piezoelectric layer to enforce the single-domain limit, i.e., the factor $A\,d$ eventually decreases. Thus even if we increase the voltage $V$ for a thicker device to keep a constant electric field (and thus a constant stress) across the device, the ``$CV^2$'' energy dissipation eventually decreases. 

The rest of the paper is organized as follows. Section~\ref{sec:model} describes the model utilized to perform the simulations. Stochastic Landau-Lifshitz-Gilbert (LLG) equation in the presence of room-temperature thermal fluctuations is solved to trace the magnetization dynamics and calculate the associated performance metrics, i.e., switching delay and energy dissipation. Simulation results showing how the performance metrics area, switching delay, and energy dissipation vary with the decrease in lateral dimensions (and the increase in thickness) of the nanomagnet are presented in Section~\ref{sec:results}. Finally, Section~\ref{sec:conclusions} discusses on the results presented and concludes the paper.

\section{Model}
\label{sec:model}

We model the piezoelectric and magnetostrictive layers as elliptical cylinders as shown in Fig.~\ref{fig:multiferroic_heterostructure}. The magnetostrictive nanomagnet lies on the $y$-$z$ plane; the major axis is aligned along the $z$-direction and the minor axis along the $y$-direction. The dimensions of the major axis, the minor axis, and the thickness are $a$, $b$, and $l$, respectively. So the nanomagnet's aspect ratio is $a/b$,  cross-sectional area is $A=(\pi/4)ab$, and volume is $\Omega=(\pi/4)abl$. The $z$-axis is the easy axis, the $y$-axis is the in-plane hard axis, and the $x$-axis is the out-of-plane hard axis. Since $l < b$, the out-of-plane hard axis is much harder than the in-plane hard axis. As $l$ becomes higher and the lateral dimensions $a$ and $b$ become smaller, it is easier for magnetization to deflect out-of-plane. In standard spherical coordinate system, $\theta$ is the polar angle and $\phi$ is the azimuthal angle of the magnetization vector. Note that when $\phi$ = $\pm90^{\circ}$, the magnetization vector lies on the plane of the nanomagnet. Any deviation from $\phi$ = $\pm90^{\circ}$ corresponds to out-of-plane excursion.

We can write the total energy of the magnetostrictive polycrystalline single-domain nanomagnet when it is subjected to uniaxial stress along the easy axis (major axis of the ellipse) as the sum of the shape-anisotropy energy and the stress-anisotropy energy~\cite{RefWorks:157,roy11,roy11_6} as 

\begin{equation}
E = B(\phi)\, sin^2\theta
\label{eq:energy}
\end{equation}
where
\begin{subequations}
\begin{align}
B(\phi) &= B_{shape}(\phi) + B_{stress},\displaybreak[3]\\
B_{shape}(\phi) &= (\mu_0/2) M_s^2 \Omega [(N_{d-yy}-N_{d-zz}) \\ 
								& \qquad \qquad + (N_{d-xx}-N_{d-yy})\,cos^2\phi],\\
B_{stress} 	&= (3/2) \lambda_s \sigma \Omega,
\label{eq:shape_stress}
\end{align}
\end{subequations}
$M_s$ is the saturation magnetization, $N_{d-mm}$ is the component of demagnetization factor along $m$-direction, which depends on the nanomagnet's dimensions~\cite{RefWorks:157,RefWorks:402}, $(3/2)\lambda_s$ is the magnetostrictive coefficient of the single-domain magnetostrictive nanomagnet~\cite{RefWorks:157}, and $\sigma$ is the stress on the nanomagnet. Note that the potential landscape of the magnetostrictive nanomagnet is \emph{symmetric} in space, however, the out-of-plane excursion of magnetization provides an equivalent \emph{asymmetry} to facilitate switching of magnetization towards its correct direction when it reaches the $x$-$y$ plane ($\theta=90^\circ$)~\cite{roy13_2}. Thermal fluctuations create an wide distribution of magnetization while switching starts and also the time required to reach $\theta=90^\circ$ is a distribution due to thermal fluctuations~\cite{roy13_2}. Therefore, it necessitates a sensing methodology to detect when magnetization reaches around $\theta=90^\circ$ but the sensing does not need to be very accurate as intrinsic dynamics provides some tolerance~\cite{roy13_2}. Note that the product of magnetostrictive coefficient and stress needs to be \emph{negative} in sign for stress-anisotropy to overcome the shape-anisotropy.

From~\eqref{eq:energy}, the in-plane barrier height (i.e., when $\theta=90^\circ$, $\phi=\pm90^\circ$, and $\sigma=0$) between two magnetization stable states ($\theta=0^\circ$ and $180^\circ$) can be written as 
\begin{equation}
E_{barrier} = (\mu_0/2) M_s^2 \Omega (N_{d-yy}-N_{d-zz}).
\label{eq:barrier_in_plane}
\end{equation}

The probability of spontaneous reversal of magnetization due to thermal fluctuations is $exp[-E_{barrier}/kT]$ according to the Boltzmann distribution, where $k$ is the Boltzmann constant, and $T$ is temperature. The minimum stress required to overcome the barrier height $E_{barrier}$ can be determined as 

\begin{equation}
\sigma_{min} = \cfrac{(\mu_0/2) M_s^2  (N_{d-yy}-N_{d-zz})}{(3/2) \lambda_s}.
\label{eq:stress_minimum}
\end{equation}

The minimum stress $\sigma_{min}$ does not directly depend on volume $\Omega$ since both shape-anisotropy and stress-anisotropy energies are proportional to volume, however, $\sigma_{min}$ depends on the particular dimensions of the nanomagnet (i.e., $a$, $b$, and $l$) through the dependence of demagnetization factors $N_{d-yy}$ and $N_{d-zz}$. It should be emphasized that the barrier height $E_{barrier}$ depends on volume $\Omega$ and to satisfy single-domain limit~\cite{RefWorks:402}, there are certain dimensions of the nanomagnet that we can choose.

The magnetization \textbf{M} of the nanomagnet has a constant magnitude  but a variable direction, so that we can represent it by a vector of unit norm $\mathbf{n_m} =\mathbf{M}/|\mathbf{M}| = \mathbf{\hat{e}_r}$ where $\mathbf{\hat{e}_r}$ is the unit vector in the radial direction in spherical coordinate system represented by ($r$,$\theta$,$\phi$).  The torque, $\mathbf{T_E}$ acting on the magnetization due to shape and stress-anisotropy is derived from the derivative of potential energy as in~\eqref{eq:energy}~\cite{roy11,roy11_6}. There is an additional torque, $\mathbf{T_{TH}}$ due to room-temperature (300 K) thermal fluctuations~\cite{roy11_6}.

\begin{figure*}[htbp]
\centering
\subfigure[]{\includegraphics[width=2.3in]{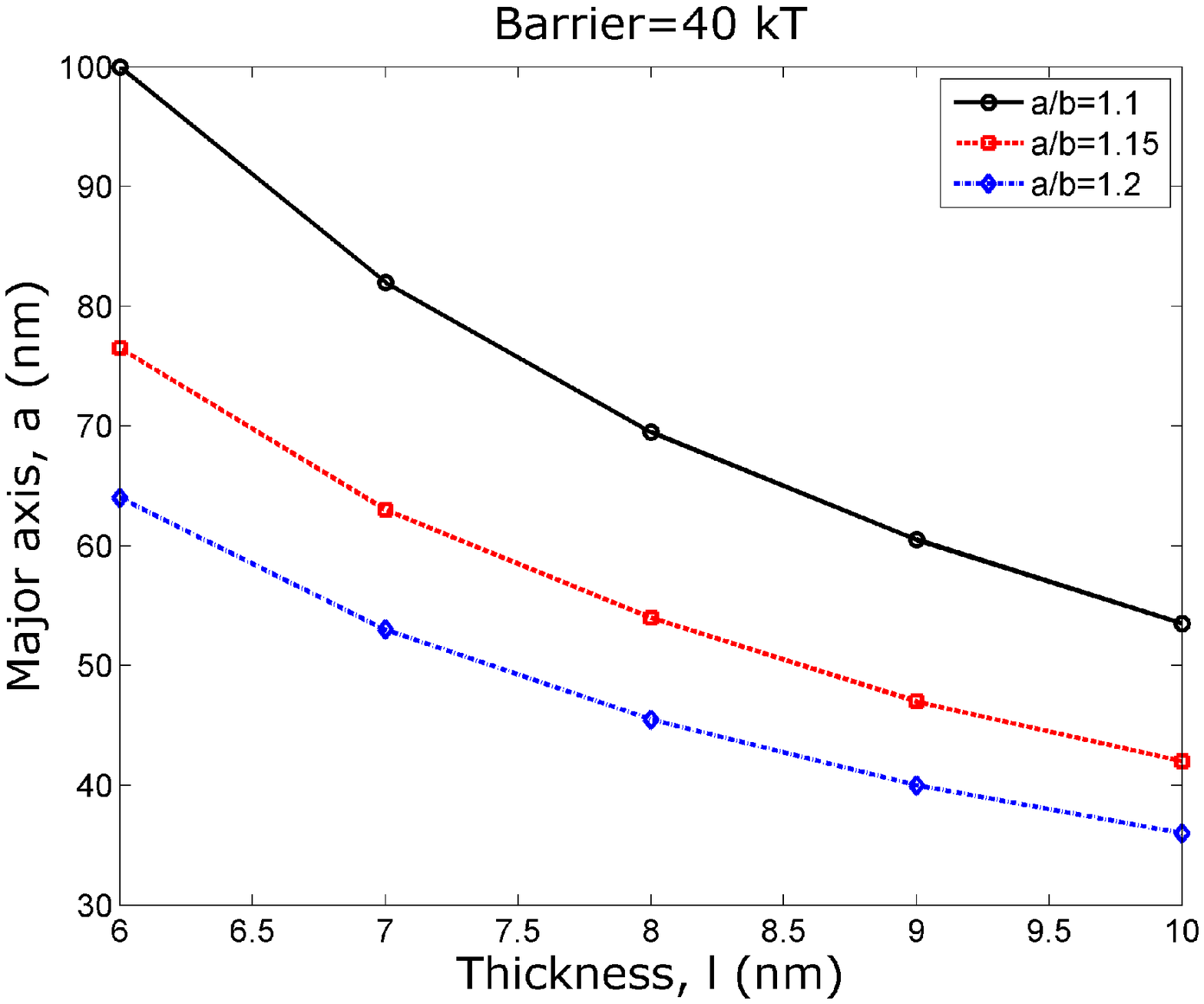}%
\label{fig:scaling_a_vs_t_abyb}}
\hfil
\subfigure[]{\includegraphics[width=2.3in]{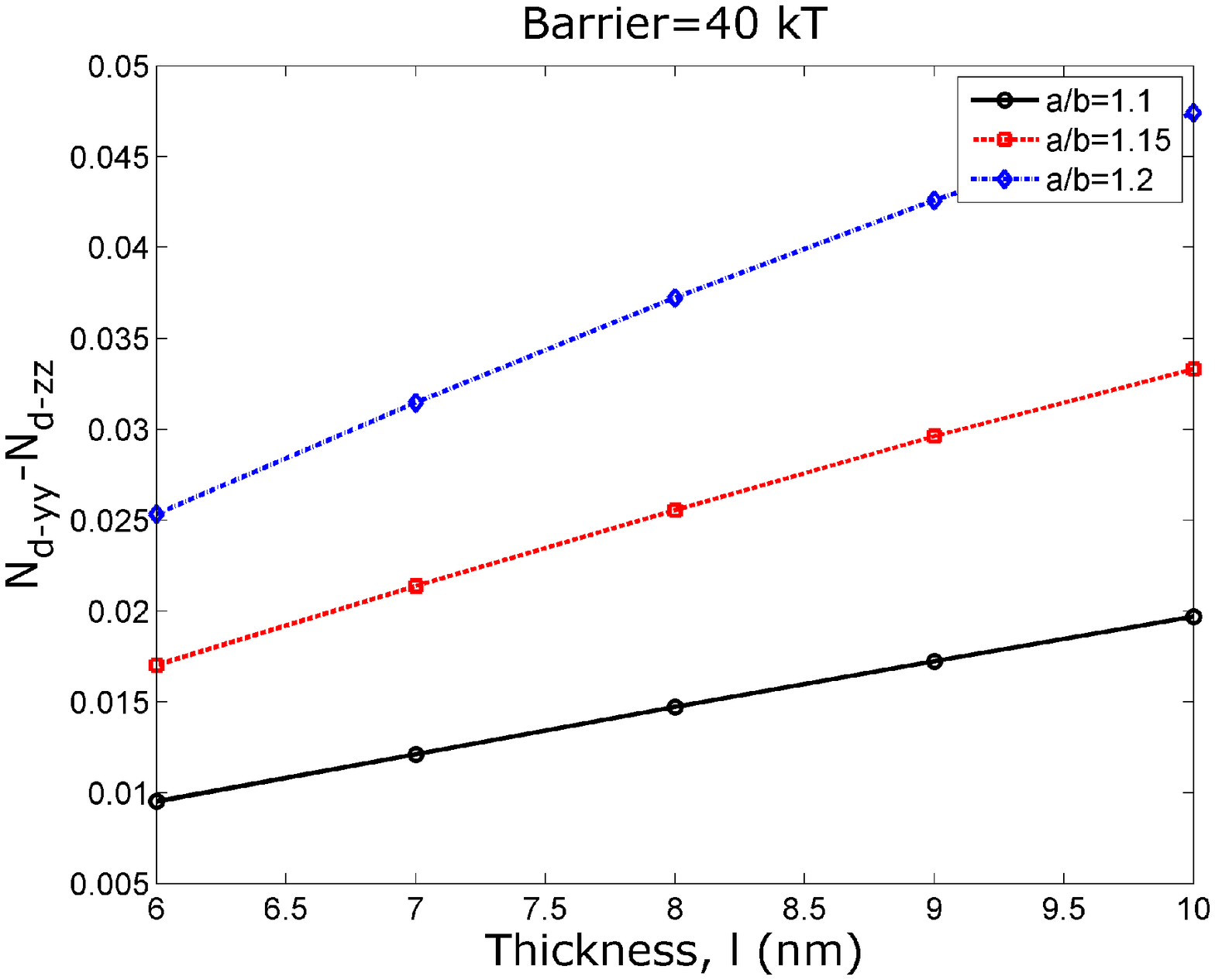}%
\label{fig:scaling_Ndyz_vs_t_abyb}}
\hfil
\subfigure[]{\includegraphics[width=2.3in]{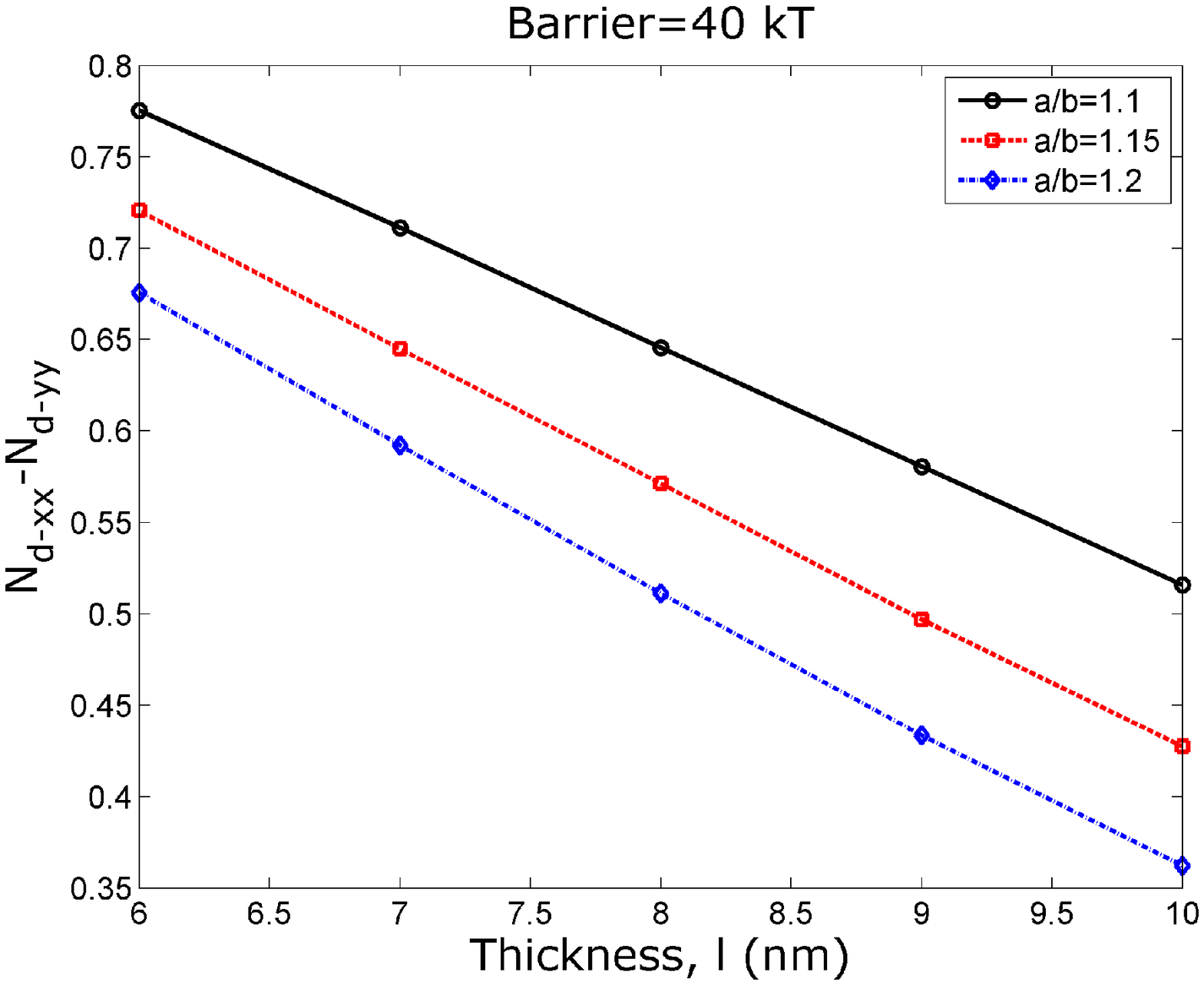}%
\label{fig:scaling_Ndxy_vs_t_abyb}}
\caption{Major axis $a$ of the nanomagnet, $N_{d-yy}-N_{d-zz}$, and $N_{d-xx}-N_{d-yy}$ versus thickness $l$ with aspect ratio $a/b$ as a parameter to satisfy the single-domain limit for an energy barrier height of 40 $kT$ at room-temperature.
(a) Major axis $a$ versus thickness $l$ with aspect ratio $a/b$ as a parameter. The major axis can be scaled down with the increase of both thickness $l$ and aspect ratio $a/b$. 
(b) $N_{d-yy}-N_{d-zz}$ versus thickness $l$ with aspect ratio $a/b$ as a parameter. $N_{d-yy}-N_{d-zz}$ increases with both thickness $l$ and aspect ratio $a/b$. 
(c) $N_{d-xx}-N_{d-yy}$ versus thickness $l$ with aspect ratio $a/b$ as a parameter. $N_{d-xx}-N_{d-yy}$ decreases with the increase of both thickness $l$ and aspect ratio $a/b$. 
}
\label{fig:scaling_a_Nd}
\end{figure*}

The magnetization dynamics under the two aforesaid torques is described by the stochastic Landau-Lifshitz-Gilbert (LLG) equation~\cite{RefWorks:162,RefWorks:161,RefWorks:186} as 
\begin{equation}
\cfrac{d\mathbf{n_m}}{dt} - \alpha \left(\mathbf{n_m} \times \cfrac{d\mathbf{n_m}}{dt} \right)\\
 = -\cfrac{|\gamma|}{M} \left\lbrack \mathbf{T_E} +  \mathbf{T_{TH}}\right\rbrack
\end{equation}
where $\alpha$ is the phenomenological damping parameter, $\gamma$ is the gyromagnetic ratio for electrons, and $M= \mu_0 M_s \Omega$.

After solving the LLG equation, we get the following coupled equations for the dynamics of $\theta$ and $\phi$~\cite{roy11_6}
\begin{multline}
\left(1+\alpha^2 \right) \cfrac{d\theta}{dt} = \frac{|\gamma|}{M} \lbrack B_{shape,\phi}(\phi) sin\theta  - 2\alpha B(\phi) sin\theta \, cos\theta \\ + \left(\alpha P_\theta + P_\phi\right)\rbrack,
\label{eq:theta_dynamics}
\end{multline}
\begin{multline}
\left(1+\alpha^2 \right) \cfrac{d \phi}{dt} = \frac{|\gamma|}{M} \lbrack \alpha B_{shape,\phi}(\phi) + 2 B(\phi) cos\theta \\
- \{sin\theta\}^{-1} \left(P_\theta - \alpha P_\phi \right) \rbrack \qquad (sin\theta \neq 0),
	\label{eq:phi_dynamics}
\end{multline}
where
\begin{subequations}
\begin{align}
B_{shape,\phi}(\phi) &= (\mu_0/2) \, M_s^2 \Omega (N_{d-xx}-N_{d-yy}) sin(2\phi),\\
P_\theta &= M\left\lbrack h_x\,cos\theta\,cos\phi + h_y\,cos\theta sin\phi - h_z\,sin\theta \right\rbrack,\\
P_\phi &= M\left\lbrack h_y\,cos\phi -h_x\,sin\phi \right\rbrack,\\
h_i &= \sqrt{\frac{2 \alpha kT}{|\gamma| M \Delta t}} \; G_{(0,1)} \qquad (i=x,y,z), \label{eq:thermal_h}
\end{align}
\end{subequations}
$1/\Delta t$ is proportional to the attempt frequency of the thermal field, $\Delta t$ is the simulation time-step, and $G_{(0,1)}$ is a Gaussian distribution with zero mean and unit variance~\cite{RefWorks:388}.

\begin{figure*}[htbp]
\centering
\subfigure[]{\includegraphics[width=2.3in]{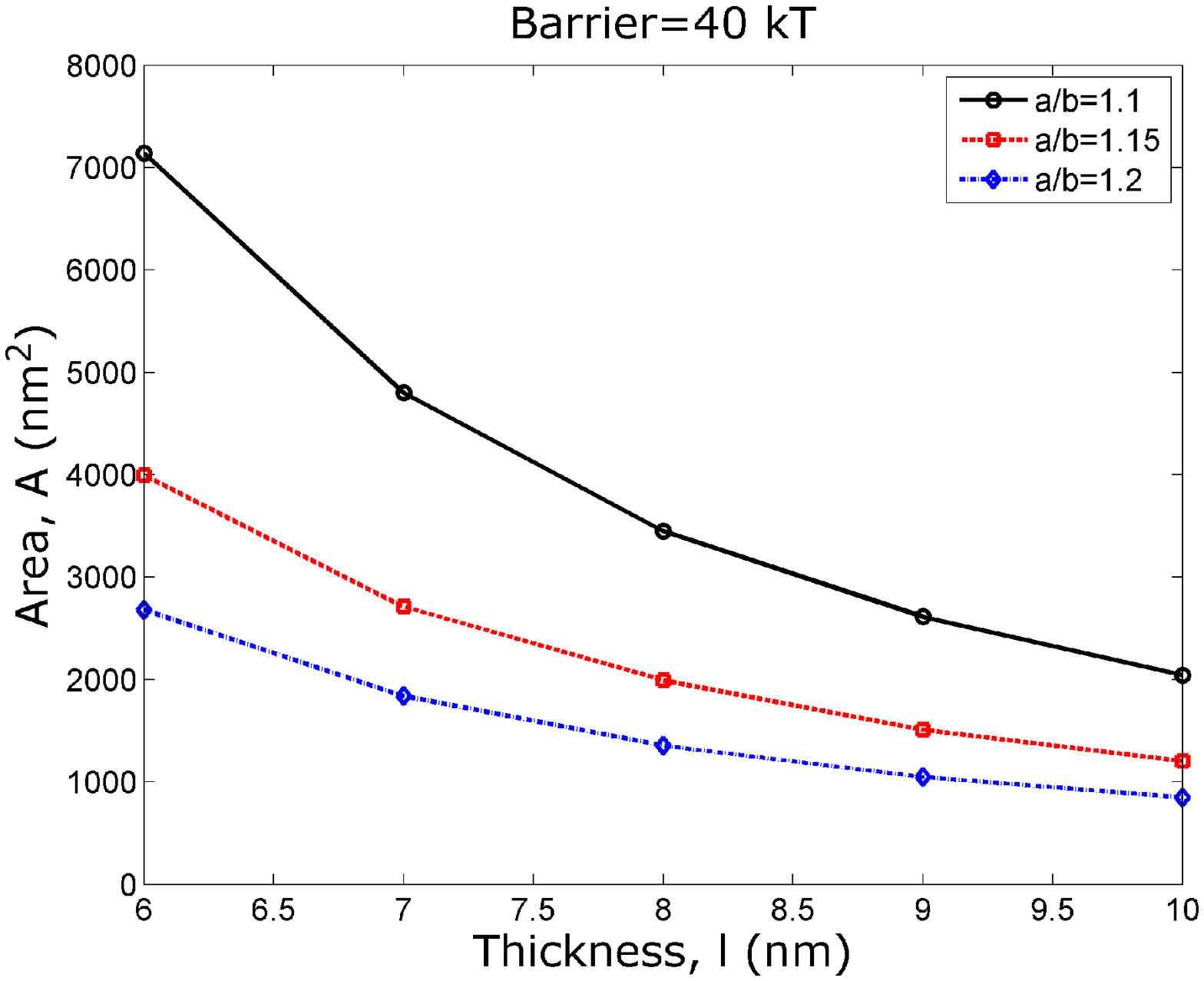}%
\label{fig:scaling_area_vs_t_abyb}}
\hfil
\subfigure[]{\includegraphics[width=2.3in]{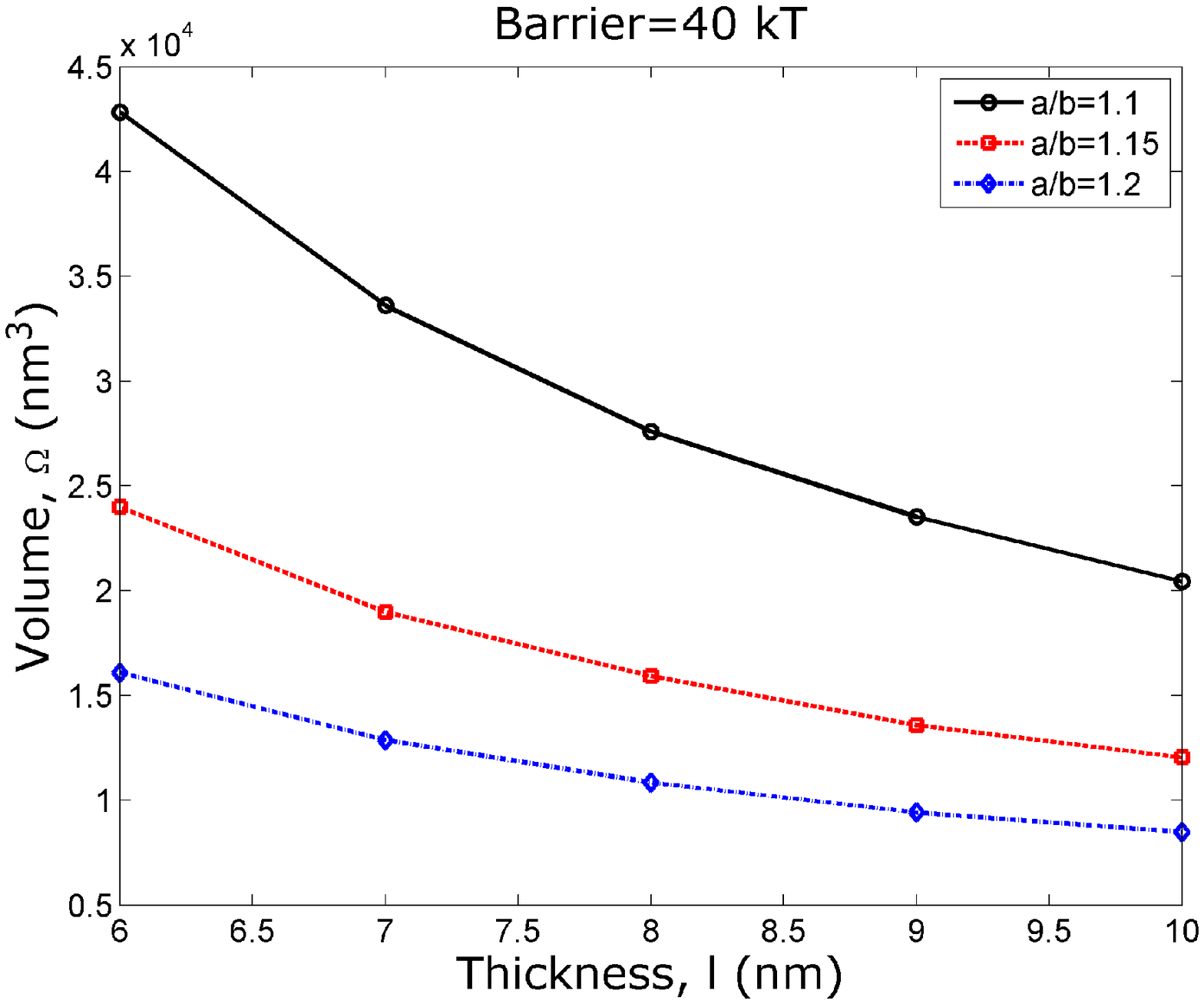}%
\label{fig:scaling_vol_vs_t_abyb}}
\hfil
\subfigure[]{\includegraphics[width=2.3in]{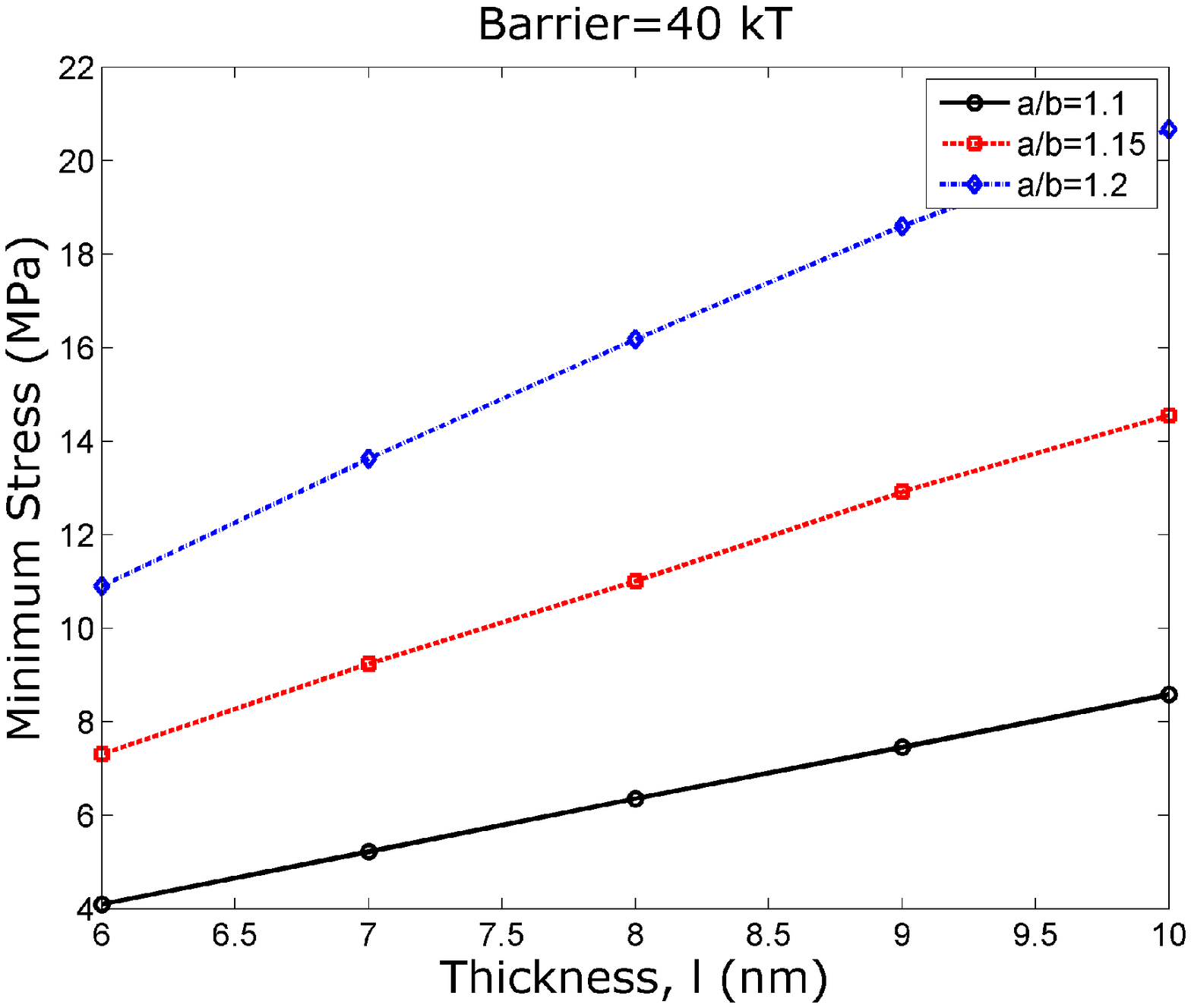}%
\label{fig:scaling_minStress_vs_t_abyb}}
\caption{Elliptical cross-sectional area $A$ of the nanomagnet, volume $\Omega$ of the nanomagnet, and the minimum stress $\sigma_{min}$ (to overcome the energy barrier height of 40 $kT$ at room-temperature) versus thickness $l$ with aspect ratio $a/b$ as a parameter.
(a) Elliptical cross sectional area $A$ of the nanomagnet versus thickness $l$ with aspect ratio $a/b$ as a parameter. Area can be scaled down with the increase of both thickness $l$ and aspect ratio $a/b$. 
(b) Volume $\Omega$ of the nanomagnet versus thickness $l$ with aspect ratio $a/b$ as a parameter. Volume eventually scales down (note that area $A$ scales down but thickness is scaled up) with the increase of both thickness $l$ and aspect ratio $a/b$. 
(c) Minimum stress [$\sigma_{min}$, see~\eqref{eq:stress_minimum}] versus thickness $l$ with aspect ratio $a/b$ as a parameter. Note that $\sigma_{min}$ is proportional to $N_{d-yy}-N_{d-zz}$ and thus the curves follow the same trend as in the Fig.~\ref{fig:scaling_Ndyz_vs_t_abyb}.
}
\label{fig:scaling_area_vol_minStress}
\end{figure*}

\section{Simulation Results}
\label{sec:results}

We consider the magnetostrictive layer to be made of polycrystalline Terfenol-D, which has the following material properties -- magnetostrictive coefficient ($(3/2)\lambda_s$): +90$\times$10$^{-5}$, Young's modulus (Y): 80 GPa, saturation magnetization ($M_s$):  8$\times$10$^5$ A/m, and Gilbert's damping constant ($\alpha$): 0.1 (Refs.~\cite{RefWorks:179,RefWorks:176,RefWorks:178,materials}). We use Terfenol-D (TbDyFe), a specially designed composite material rather than common magnetic materials (e.g., iron, nickel, or cobalt) since Terfenol-D has 30 times higher magnetostrictive coefficient in magnitude~\cite{roy11}. Another good choice may be FeGa alloy, which has magnetostrictive coefficient +15$\times$10$^{-5}$~\cite{RefWorks:167}. Since Terfenol-D has \emph{positive} magnetostrictive coefficient, a \emph{compressive} stress is required for magnetization to switch~\cite{roy13_spin}. In our convention, a \emph{tensile} stress is positive and a \emph{compressive} stress is negative. 

The piezoelectric layer is considered to be made of lead-zirconate-titanate (PZT), which has a dielectric constant of 1000~\cite{roy13_spin}. The PZT layer is assumed to be four times thicker than the magnetostrictive layer so that any strain generated in it is transferred almost completely to the magnetostrictive layer~\cite{roy13_spin}. The maximum strain on the PZT layer is considered to be 500 ppm~\cite{RefWorks:170,RefWorks:563} and it would require an electric field of 2.78 MV/m because $d_{31}$=1.8$\times$10$^{-10}$ m/V for PZT~\cite{pzt2}. (We can also use relaxor ferroelectrics for the piezoelectric layer e.g., PMN-PT, PZN-PT that have high piezoelectric coefficients and generate anisotropic strain, hence it can decrease the electric field further~\cite{pmnpt,RefWorks:806,RefWorks:790}.) The corresponding stress in Terfenol-D is the product of the generated strain ($500\times10^{-6}$) and the Young's modulus (80 GPa). Hence, 40 MPa is the maximum stress that can be generated in the Terfenol-D nanomagnet. For example, to generate 30 MPa stress, it would require voltages of 83.3 mV and 50 mV for nanomagnets with thicknesses 10 nm and 6 nm, respectively.


We always ensure that the magnetostrictive nanomagnet has a single ferromagnetic domain~\cite{RefWorks:402,RefWorks:133} for varied thickness and aspect ratio. The performance metrics switching delay and energy dissipation are determined by solving stochastic Landau-Lifshitz-Gilbert equation in the presence of room-temperature (300 K) thermal fluctuations following the prescription in Ref.~\cite{roy11_6,roy11_2}. We assume a 50 ps ramp for stress on the magnetostrictive nanomagnet~\cite{roy11_6}.

\subsection{Lateral Dimensions and Demagnetization Factors}

Fig.~\ref{fig:scaling_a_vs_t_abyb} plots the major axis $a$ of the nanomagnet versus its thickness $l$ with aspect ratio $a/b$ as a parameter for an energy barrier height of 40 $kT$ at room-temperature. The decrease in lateral dimensions with the increase in thickness are adjusted to satisfy the single-domain limit assumption~\cite{RefWorks:402}. With these dimensions, the demagnetization factors $N_{d-yy}$ and $N_{d-zz}$ both increase with thickness and aspect ratio, however, $N_{d-xx}$ has the opposite trend~\cite{RefWorks:402}. Fig.~\ref{fig:scaling_Ndyz_vs_t_abyb} plots the difference $N_{d-yy}-N_{d-zz}$, which increases with both thickness and aspect ratio. Fig.~\ref{fig:scaling_Ndxy_vs_t_abyb} plots $N_{d-xx}-N_{d-yy}$, which decreases with the increase of both thickness and aspect ratio.

\subsection{Critical Stress and Success Rate of Switching}

Figs.~\ref{fig:scaling_area_vs_t_abyb} and~\ref{fig:scaling_vol_vs_t_abyb} show the trend in lateral area and volume of the nanomagnet while varying its thickness with aspect ratio as a parameter. These plots can be derived from the Fig.~\ref{fig:scaling_a_vs_t_abyb}. The decrease in lateral area bodes well with the requirement of scaling down the device size to enhance device density on a chip. However, the decrease in volume has to be compensated by increasing the in-plane shape-anisotropy ($\propto$ $N_{d-yy}-N_{d-zz}$) to keep the energy barrier height same as 40 $kT$ (see~\eqref{eq:barrier_in_plane}). Thus, the curves in Fig.~\ref{fig:scaling_Ndyz_vs_t_abyb} show the opposite trend to that of Fig.~\ref{fig:scaling_vol_vs_t_abyb} with the thickness and aspect ratio. 

Fig.~\ref{fig:scaling_minStress_vs_t_abyb} shows the minimum stress $\sigma_{min}$ required for stress-anisotropy to overcome the shape-anisotropy energy barrier of the nanomagnet. Since both the anisotropies are proportional to volume, $\sigma_{min}$ is independent of volume but it is proportional to $N_{d-yy}-N_{d-zz}$ for a fixed barrier height of 40 $kT$ (see~\eqref{eq:stress_minimum} and Fig.~\ref{fig:scaling_Ndyz_vs_t_abyb}).

\begin{figure}[bthp]
\centering
\includegraphics[width=2.3in]{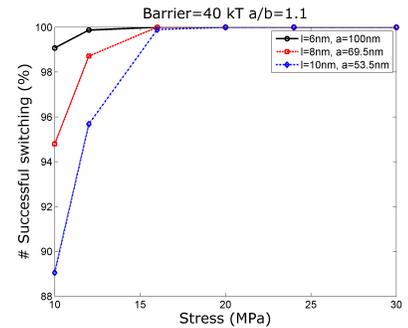}%
\caption{Percentage of successful switching events ($SR$) as a function of stress (10-30 MPa) versus thickness $l$ with aspect ratio $a/b$ as a parameter for an energy barrier height of 40 $kT$. A moderately large number of simulations (10000) are performed in the presence of room-temperature (300 K) thermal fluctuations to generate each data point in these plots. At lower stress-levels (10-15 MPa), the number of successful switching events decreases with increasing thickness.}
\label{fig:scaling_success_vs_stress_t_a}
\end{figure}

Fig.~\ref{fig:scaling_success_vs_stress_t_a} plots the successful switching rate versus stress for different thicknesses of the nanomagnet. A higher stress keeps the magnetization more out-of-plane ($x$-direction) and thus it is conducive to a higher success rate of switching~\cite{roy13_2}. For a fixed aspect ratio, a higher thickness of the nanomagnet corresponds to lower volume and thus lesser stress-anisotropy is produced at lower stress levels. This is the reason behind having lower success rate of switching with increasing thickness at the low stress levels.

\subsection{Switching Delay and Energy Dissipation}

\begin{figure}[tbp]
\centering
\subfigure[]{\includegraphics[width=2.3in]{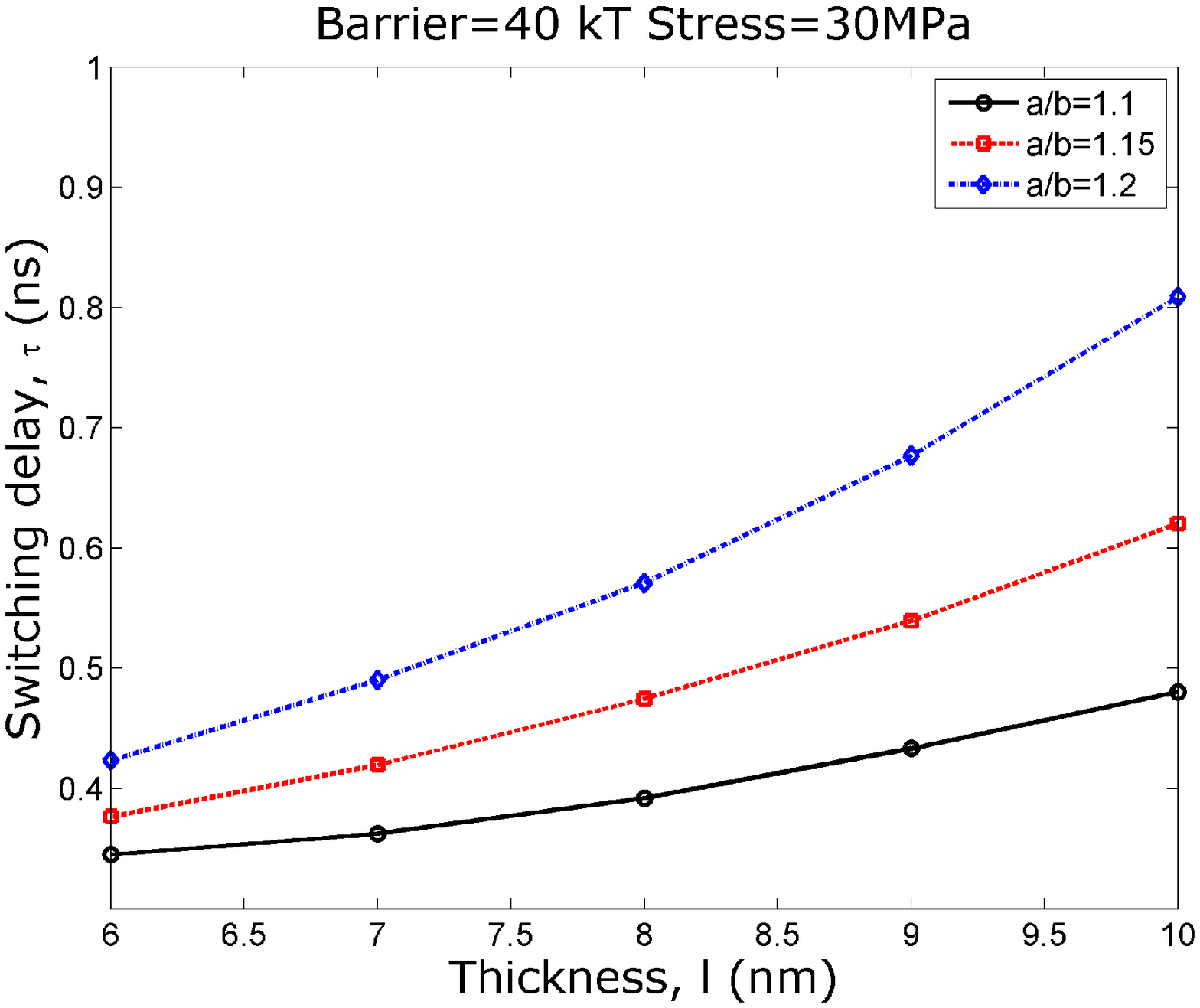}%
\label{fig:scaling_delay_vs_t_abyb}}
\hfil
\subfigure[]{\includegraphics[width=2.3in]{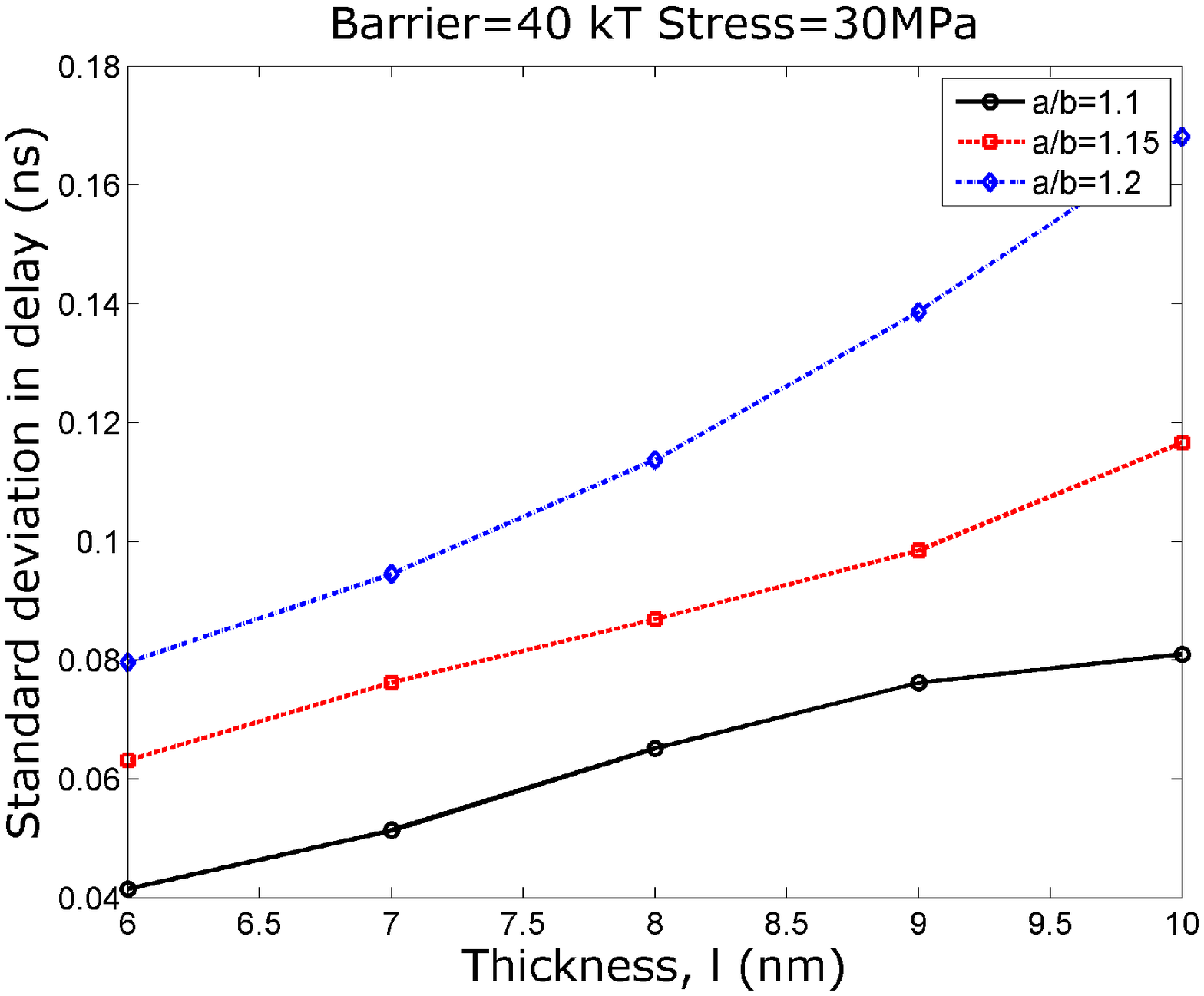}%
\label{fig:scaling_delay_std_vs_t_abyb}}
\caption{Mean and standard deviation of switching delay ($\tau_{mean}$ and $\tau_{std}$, respectively) versus thickness $l$ with aspect ratio $a/b$ as a parameter for an energy barrier height of 40 $kT$ at room-temperature and stress 30 MPa. Each data point in these plots is calculated from 10000 simulations in the presence of room-temperature thermal fluctuations.
(a) Thermal mean switching delay versus thickness $l$ with aspect ratio $a/b$ as a parameter. The mean switching delay increases with both thickness (for a fixed aspect ratio) and aspect ratio (for a fixed thickness).
(b) Standard deviation in switching delay versus thickness $l$ with aspect ratio $a/b$ as a parameter. The standard deviation increases with both thickness (for a fixed aspect ratio) and aspect ratio (for a fixed thickness).
}
\label{fig:scaling_delay}
\end{figure}

\begin{figure}[tbp]
\centering
\subfigure[]{\includegraphics[width=2.3in]{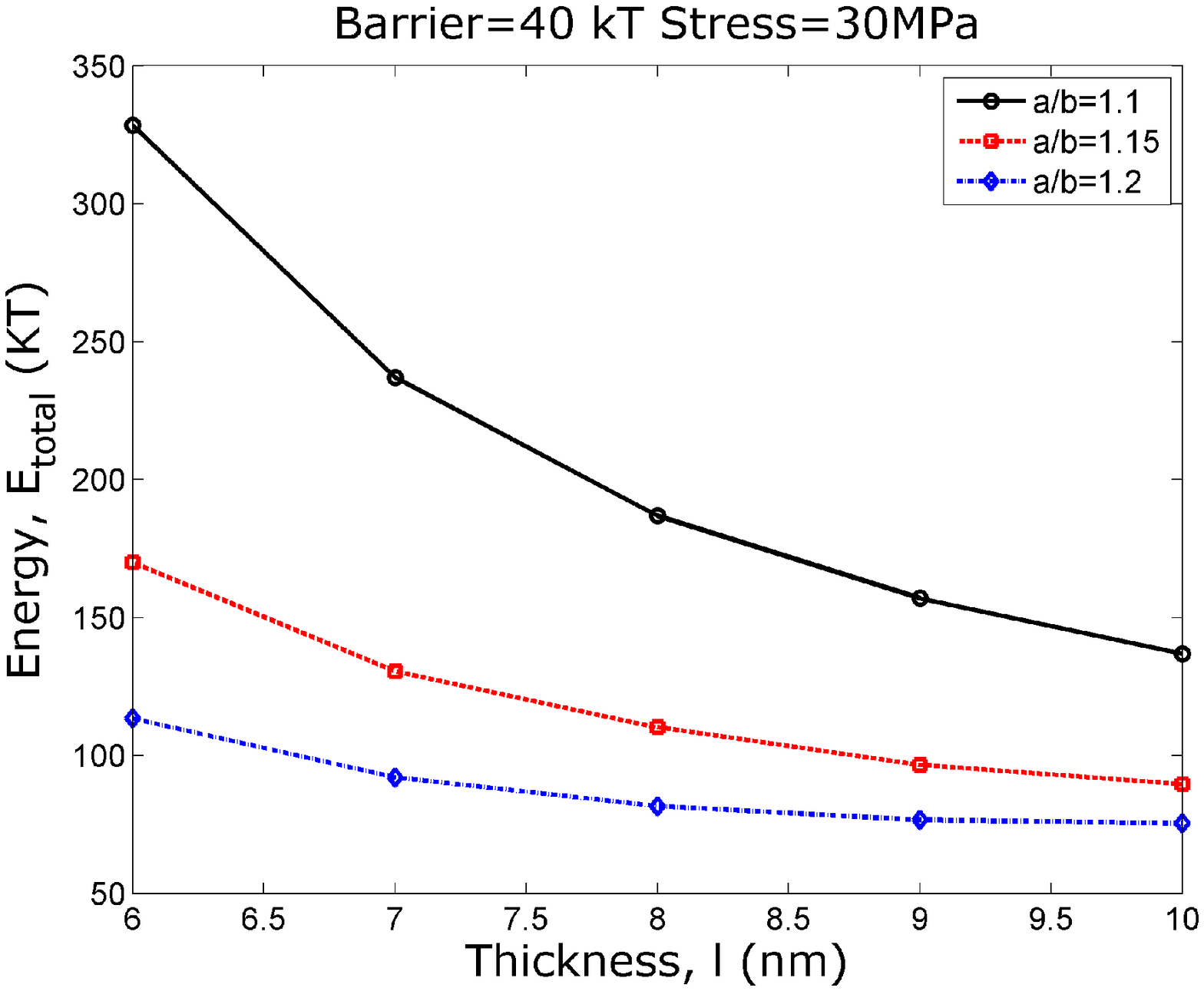}%
\label{fig:scaling_energy_vs_t_abyb}}
\hfil
\subfigure[]{\includegraphics[width=2.3in]{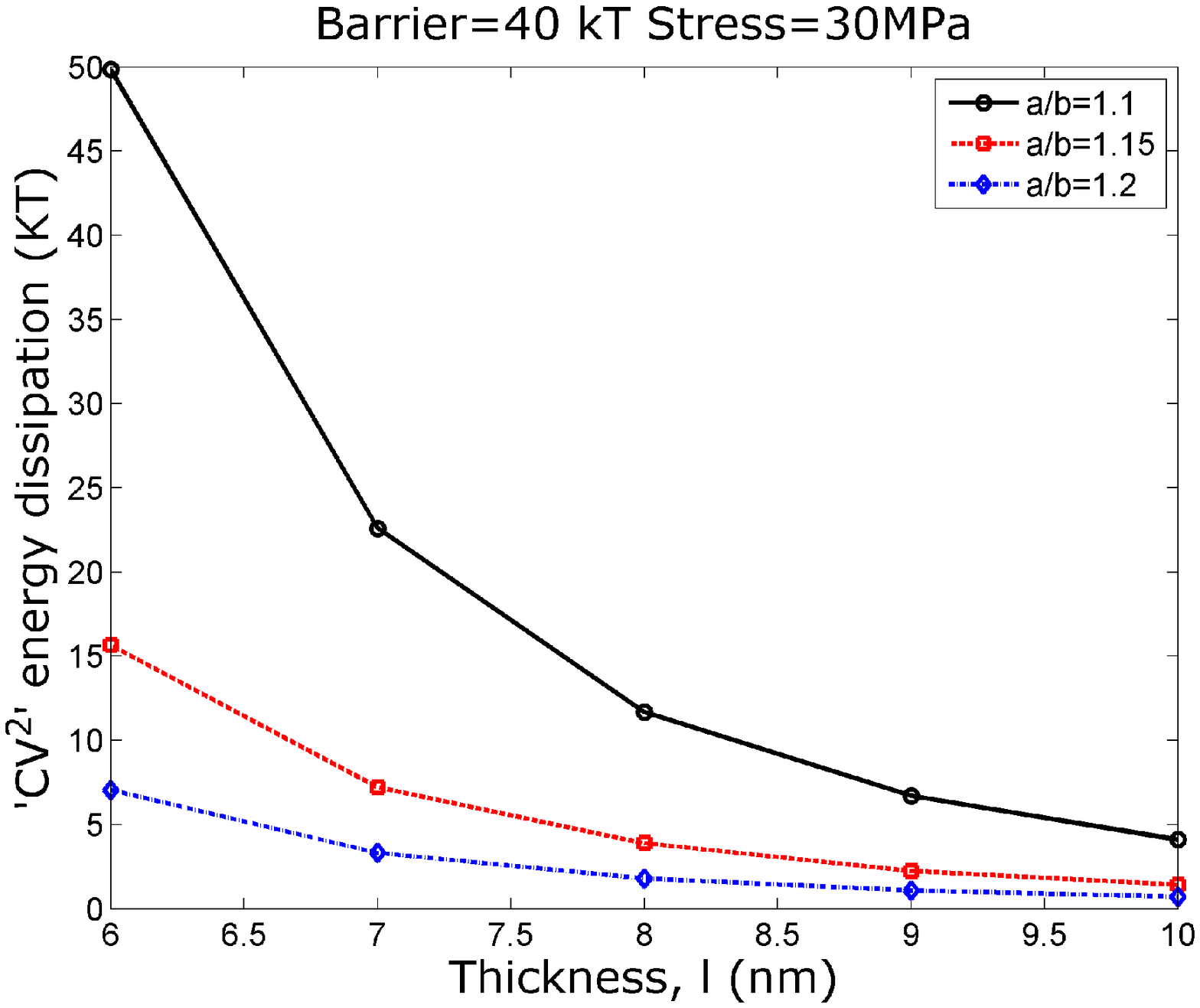}%
\label{fig:scaling_energyCV2_vs_t_abyb}}
\caption{Total and ``$CV^2$'' energy dissipation ($E_{total}$ and $E_{`CV^2\prime}$, respectively) versus thickness $l$ with aspect ratio $a/b$ as a parameter for an energy barrier height of 40 $kT$ at room-temperature and stress 30 MPa. Each data point in these plots is calculated from 10000 simulations in the presence of room-temperature thermal fluctuations.
(a) Total energy dissipation versus thickness $l$ with aspect ratio $a/b$ as a parameter. Total energy dissipation increases with both thickness (for a fixed aspect ratio) and aspect ratio (for a fixed thickness).
(b) ``$CV^2$'' energy dissipation versus thickness $l$ with aspect ratio $a/b$ as a parameter. ``$CV^2$'' energy dissipation decreases with the increase of both thickness (for a fixed aspect ratio) and aspect ratio (for a fixed thickness).
}
\label{fig:scaling_energy}
\end{figure}

Figs.~\ref{fig:scaling_delay} and~\ref{fig:scaling_energy} plot the metrics switching delay and energy  dissipation versus thickness with aspect ratio as a parameter for stress 30 MPa and an energy barrier height of 40 $kT$ at room-temperature. In Fig.~\ref{fig:scaling_delay}, we see that both the mean and standard deviation of switching delay become higher with increasing thickness. The reason behind the increase in switching delay is that it requires a minimum stress $\sigma_{min}$ to overcome the energy barrier; $\sigma_{min}$ increases with thickness (see Fig.~\ref{fig:scaling_minStress_vs_t_abyb}) and hence for a fixed stress of 30 MPa, the torque acting on the magnetization becomes lesser with higher thicknesses. The standard deviation in switching delay increases due to the decrease of nanomagnet's volume with the increasing thickness (see Fig.~\ref{fig:scaling_vol_vs_t_abyb}), which makes magnetization more prone to thermal fluctuations (see~\eqref{eq:thermal_h}). The energy dissipation curves show an opposite trend because of the delay-energy trade-off. The ``$CV^2$'' energy dissipation as shown in Fig.~\ref{fig:scaling_energyCV2_vs_t_abyb} decreases with increasing thickness as explained in the Section~\ref{sec:introduction}.

\section{Discussions and Conclusions}
\label{sec:conclusions}

We have analyzed the area-delay-energy trade-offs of strain-coupled multiferroic composite devices. It turns out that it is possible to increase the thickness of the magnetostrictive nanomagnet while reducing the lateral area of the nanomagnet keeping the single-domain limit assumption intact. However, the resultant volume of the nanomagnet eventually decreases with the decrease of lateral area. This weakens the induced stress-anisotropy in the nanomagnet for a material with fixed magnetostrictive coefficient and a certain stress value. The switching delay increases with the increasing thickness, while energy dissipation follows the opposite trend. An increase of stress can make the switching delay lower, however, there is a limit on stress that we can apply since it is dictated by the dielectric breakdown of the piezoelectric layer and the associated reliability issues. Thus, increasing magnetostrictive coefficient by superior material synthesis may be a solution to decrease switching delay as we reduce the lateral area (and increase the thickness) of a device. For example, using a twice magnetostrictive coefficient of Terfenol-D (see Ref.~\cite{RefWorks:612}, however, this is for single-crystal case rather than polycrystalline Terfenol-D) for thickness 10 nm, we can reduce the switching delay to the value same as for thickness 6 nm, while achieving around two times reduction in energy dissipation. Both reducing lateral dimensions and increasing magnetostrictive coefficient are challenging in fabrication and material synthesis procedures, which can increase the device density on a chip.

Note that it may be possible to devise different switching methodologies with strain-mediated multiferroic composites having better performance metrics particularly reducing the lateral area of the devices further. However, since the strain generated is proportional to volume and with scaling volume reduces, it may be very useful to go beyond volumetric effect and employ surface-related phenomena for magnetoelectric coupling~\cite{RefWorks:649,roy14_2}. 

With the growing experimental efforts, such devices may be a staple of modern non-volatile logic and memory systems for our future information processing paradigm and also could be deemed suitable for energy harvesting applications. 


\appendices
\section{Fluctuations of Magnetization around Easy Axis due to Thermal Agitations}

\begin{figure}[htbp]
\centering
\includegraphics[width=2.3in]{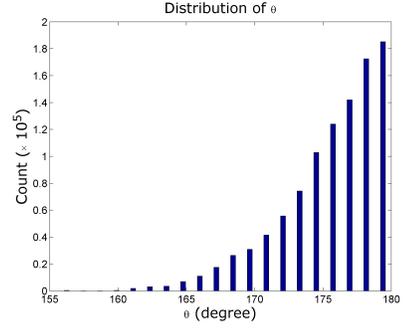}%
\caption{Distribution of polar angle $\theta$ when magnetization is fluctuating around the easy axis $\theta=180^\circ$ for an energy barrier height of 40 $kT$. This plot is particularly for a nanomagnet with 10 nm thickness and aspect ratio of 1.1. The most likely value of this distribution is $180^\circ$ but the mean of this distribution is $175.55^\circ$.}
\label{fig:theta_distribution_barrier_40KT_abyb1p1_t10nm}
\end{figure}

Magnetization in a nanomagnet cannot move exactly from the easy axes ($\theta=0^\circ, 180^\circ$) because the torque acting on magnetization become zero at these points~\cite{roy13_2}. Fortunately, thermal fluctuations can deflect magnetization from these positions. Fig.~\ref{fig:theta_distribution_barrier_40KT_abyb1p1_t10nm} shows the distribution of polar angle $\theta$ when magnetization is fluctuating around the easy axis $\theta=180^\circ$ for an energy barrier height of 40 $kT$ and for a nanomagnet with 10 nm thickness and aspect ratio of 1.1. Similar Boltzmann distributions are achieved for different dimensions of the nanomagnet but for same barrier height 40 $kT$. The mean of this distribution is used as the magnetization's initial polar angle to start with. 

\begin{figure}[htbp]
\centering
\subfigure[]{\includegraphics[width=2.3in]{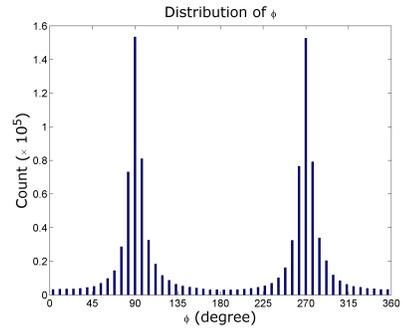}%
\label{fig:phi_distribution_barrier_40KT_abyb1p1_t6nm}}
\hfil
\subfigure[]{\includegraphics[width=2.3in]{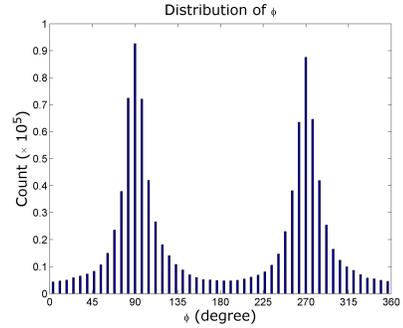}%
\label{fig:phi_distribution_barrier_40KT_abyb1p1_t10nm}}
\caption{Distributions of azimuthal angle $\phi$ when magnetization is fluctuating around the easy axis $\theta=180^\circ$ for an energy barrier height of 40 $kT$. 
(a) Distribution of azimuthal angle $\phi$ for a nanomagnet with 6 nm thickness and aspect ratio 1.1.
(b) Distribution of azimuthal angle $\phi$ for a nanomagnet with 10 nm thickness and aspect ratio 1.1.
}
\label{fig:scaling_phi_distribution}
\end{figure}

Fig.~\ref{fig:scaling_phi_distribution} shows the distribution of azimuthal angle $\phi$ when magnetization is fluctuating around the easy axis $\theta=180^\circ$ for an energy barrier height of 40 $kT$. Two cases corresponding to two different thicknesses of the nanomagnet (6 nm and 10 nm) are considered. The distributions are Gaussian peaked with the peaks on the plane of the nanomagnet ($\phi=\pm90^\circ$). We have assumed the initial value of $\phi$ as $90^\circ$, however, assuming the other in-plane angle of magnetization produces similar magnetization dynamics due to symmetry of the underlying equations. Note that the distribution corresponding to the 10 nm thickness one is less peaked since it is easier for the magnetization to deflect out-of-plane with higher thickness ($N_{d-xx}$ is decreased at higher thickness, see Fig.~\ref{fig:scaling_Ndxy_vs_t_abyb}). 

\begin{table*}[!t]
\renewcommand{\arraystretch}{1.3}
\caption{Performance metrics e.g., success rate of switching ($SR$), mean switching delay ($\tau_{mean}$), standard deviation in switching delay ($\tau_{std}$), total energy dissipation ($E_{total}$), and `$CV^2$' energy dissipation $E_{`CV^2\prime}$ for a Terfenol-D/PZT strain-mediated multiferroic composite device with respect to thickness $l$ with the aspect ratio ($a/b$) of the elliptical cross-section of the nanomagnet as a parameter. The stress is 30 MPa and the energy barrier height is 40 $kT$. The derived quantities that are given are $N_{d-yz}=N_{d-yy}-N_{d-zz}$, $N_{d-xy}=N_{d-xx}-N_{d-yy}$, Area (A), Volume ($\Omega$), and minimum stress to overcome the energy barrier of the nanomagnet $\sigma_{min}$. The performance metrics are calculated from 10000 simulations.
}
\label{tab:Table1}
\centering
\begin{tabular}{r|r|r|r|r|r|r|r|r|r|r|r|r}
\hline
\multicolumn{1}{c|}{$l$} & \multicolumn{1}{c|}{$a/b$} &  \multicolumn{1}{c|}{a}  &  \multicolumn{1}{c|}{$N_{d-yz}$} &  \multicolumn{1}{c|}{$N_{d-xy}$} &  \multicolumn{1}{c|}{$A$}  &  \multicolumn{1}{c|}{$\Omega$} &  \multicolumn{1}{c|}{$\sigma_{min}$} &  \multicolumn{1}{c|}{$SR$} &  \multicolumn{1}{c|}{$\tau_{mean}$} & \multicolumn{1}{c|}{$\tau_{std}$} & \multicolumn{1}{c|}{$E_{total}$} & \multicolumn{1}{c}{$E_{` CV^2 \prime}$}\\
\multicolumn{1}{c|}{(nm)} & \multicolumn{1}{c|}{} & \multicolumn{1}{c|}{(nm)} & \multicolumn{1}{c|}{}  & \multicolumn{1}{c|}{}  & \multicolumn{1}{c|}{(nm$^2$)} & \multicolumn{1}{c|}{(nm$^3$)} & \multicolumn{1}{c|}{(MPa)} & \multicolumn{1}{c|}{(\%)} & \multicolumn{1}{c|}{(ns)} & \multicolumn{1}{c|}{(ns)} & \multicolumn{1}{c|}{(kT)} & \multicolumn{1}{c}{(kT)}\\ 
\hline\hline
 6	& 1.10	& 100.0	& 0.0095	& 0.7754	&  7139.98	& 42839.90	&    4.25	& 100.00	& 0.35	& 0.042	& 328.45	& 49.85 \\
  	& 1.15	&  76.5	& 0.0170	& 0.7206	&  3996.82	& 23980.94	&    7.60	& 100.00	& 0.38	& 0.063	& 169.95	& 15.66 \\
  	& 1.20	&  64.0	& 0.0253	& 0.6756	&  2680.83	& 16084.95	&   11.32	& 100.00	& 0.42	& 0.080	& 113.40	&  7.05 \\
 \hline
 7	& 1.10	&  82.0	& 0.0121	& 0.7112	&  4800.92	& 33606.47	&    5.41	& 100.00	& 0.36	& 0.051	& 237.05	& 22.59 \\
  	& 1.15	&  63.0	& 0.0214	& 0.6449	&  2710.65	& 18974.54	&    9.55	& 100.00	& 0.42	& 0.076	& 130.51	&  7.21 \\
  	& 1.20	&  53.0	& 0.0315	& 0.5921	&  1838.49	& 12869.40	&   14.05	& 100.00	& 0.49	& 0.094	&  92.11	&  3.32 \\
\hline
 8	& 1.10	&  69.5	& 0.0147	& 0.6456	&  3448.79	& 27590.32	&    6.57	& 100.00	& 0.39	& 0.065	& 186.85	& 11.67 \\
  	& 1.15	&  54.0	& 0.0255	& 0.5712	&  1991.50	& 15931.97	&   11.41	& 100.00	& 0.47	& 0.087	& 110.15	&  3.89 \\
  	& 1.20	&  45.5	& 0.0372	& 0.5112	&  1354.98	& 10839.80	&   16.63	& 100.00	& 0.57	& 0.114	&  81.65	&  1.80 \\
\hline
 9	& 1.10	&  60.5	& 0.0172	& 0.5806	&  2613.41	& 23520.71	&    7.70	& 100.00	& 0.43	& 0.076	& 157.00	&  6.71 \\
  	& 1.15	&  47.0	& 0.0296	& 0.4968	&  1508.65	& 13577.83	&   13.23	& 100.00	& 0.54	& 0.098	&  96.55	&  2.24 \\
  	& 1.20	&  40.0	& 0.0426	& 0.4335	&  1047.20	& 9424.78	&   19.04	&  99.99	& 0.68	& 0.139	&  76.61	&  1.08 \\
\hline
10	& 1.10	&  53.5	& 0.0197	& 0.5159	&  2043.64	& 20436.42	&    8.79	& 100.00	& 0.48	& 0.081	& 136.85	&  4.10 \\
  	& 1.15	&  42.0	& 0.0333	& 0.4275	&  1204.73	& 12047.32	&   14.88	& 100.00	& 0.62	& 0.117	&  89.59	&  1.43 \\
  	& 1.20	&  36.0	& 0.0474	& 0.3620	&   848.23	& 8482.30	&   21.18	&  99.91	& 0.81	& 0.168	&  75.44	&  0.71 \\
\hline
\end{tabular}
\end{table*}

\section{Results in Tabular Format}

Table~\ref{tab:Table1} tabulates the simulation results presented in this paper. Different performance metrics e.g., successful switching rate, switching delay, and energy dissipation are given as a function of thickness with aspect ratio of the elliptical cross-section of the nanomagnet as a parameter. As we scale down the lateral dimensions of the nanomagnet, we can increase the thickness to some extent to keep the single-domain assumption valid. However, the volume still decreases with scaling and the volumetric strain decreases proportionately. Therefore the minimum stress $\sigma_{min}$ required to topple the energy barrier of the nanomagnet increases. Switching delay (both mean and standard deviation) increases and energy dissipation (both $E_{total}$ and `$CV^2$' one) decreases as already discussed in the paper. 

For a given thickness, with the increase of aspect ratio, the anisotropy increases and it allows us to decrease the lateral area so that the energy barrier height can be kept constant. The volume decreases in turn and the minimum stress $\sigma_{min}$ increases. Switching delay (both mean and standard deviation) increases and energy dissipation (both $E_{total}$ and `$CV^2$' one) decreases as already discussed in the paper. For thicknesses of 9 and 10 nm, and aspect ratio of 1.2, we notice some switching failures among the 10000 simulations due the increase in the minimum stress $\sigma_{min}$. Note that although the stress on the nanomagnet 30 MPa is higher than $\sigma_{min}$, sometimes thermal fluctuations can scuttle the magnetization causing switching failures.


\end{document}